\documentclass{article}

\usepackage{graphicx}  
\usepackage{amsmath}   
\usepackage[compress]{cite}
\usepackage{amssymb}   
\usepackage{bm} 

\usepackage{mathrsfs}

\usepackage[active]{srcltx}

\def\eq#1{{Eq.~(\ref{#1})}}

\def\bk#1#2#3{{\langle #1|#2|#3\rangle}}  

\title{Path integrals for the relativistic particle: Some conceptual and pedagogical comments}

\author{T. Padmanabhan\\
IUCAA, Post Bag 4, Ganeshkhind,
 Pune - 411 007, India.\\
email: paddy@iucaa.in}

\date{ }

\begin{document}

\maketitle

\begin{abstract}
In my textbook on Quantum Field Theory \cite{tpqft} and in a recent paper \cite{tpejc2018}, I advocated a  lattice regularization procedure  for defining the path integral for the relativistic particle, using the non-quadratic action containing a square root. I also provided an \textit{interpretation}  of this result in terms of the Jacobi action principle. This note clarifies several conceptual and pedagogical issues related to this approach and highlights some interesting open questions which this result leads to.
\end{abstract}

\section{Introduction}

The classical dynamics of both non-relativistic and  relativistic particles can be obtained from  extremum principles based on action functionals. (The action functional is not unique and many functionals will lead to the same dynamical equations under appropriate variation.) It is also possible to obtain the quantum mechanics of the non-relativistic particle from a path integral using the primary amplitude $\exp(iA)$ where $A$ is  the appropriate  action functional. 
This path integral can be defined so that it leads to the propagator $G_{\rm NR}(x_2,x_1)$ which, in turn, will evolve the single-particle wave function from $t=t_1$ to $t=t_2$, thereby describing the quantum dynamics of a single particle.\footnote{\textit{Notations:} I use the signature which is mostly positive and work in 1+3 dimensions. Greek superscripts etc. range over 1, 2, 3 and Latin superscripts etc. range over $0 - 3$. The square bracket, like in $A_1 [x^\alpha(t)]$, denotes a functional when appropriate. Summation convention is assumed throughout. Events in spacetime are often denoted by just $x_1, x_2, x ....$ etc without superscripts, when no confusion is likely to arise. I use units with $c=\hbar = 1$.}

Can one, similarly, define a \textit{relativistic} path integral using the relativistic action functional and obtain a theory which incorporates both QM and SR? From a practical point of view this is not a useful procedure. Particles are best described as excitations of underlying fields which carry representations of the Lorentz group.  This is the standard procedure which allows us to combine SR and QM, especially when you want to describe particles with spin or interactions. (In fact, this is possibly why the standard textbooks in QFT --- except mine \cite{tpqft}! --- do not discuss the relativistic path integral.)   QFT also tells us that  combining SR with QM is not consistent with a single particle description --- which remained valid \textit{both} in classical theory as well as in NRQM --- so it is not a priori obvious what the relativistic propagator, defined by a path integral, describes. 

Nevertheless, there are certain advantages in studying a path integral based on relativistic single particle actions. One would expect that, at least in the case of spin-less particle, such a formulation should exist and lead to the relativistic (Feynman) propagator $G_R(x_2,x_1)$. \textit{We will see that this is indeed the case.} The importance of this approach lies in the fact that it raises several fundamental questions which are not apparent in the standard formulation of QFT. I have explored some of these aspects in my textbook on QFT \cite{tpqft} as well as in a recent detailed paper \cite{tpejc2018}. 
In this note, I discuss several conceptual and pedagogical aspects of this problem using the Ref. \cite{tpejc2018} as the backdrop.\footnote{Since many of the mathematical and algebraic details are already covered in \cite{tpejc2018}, I will not repeat them in this note and instead concentrate on conceptual inter-relationships. The reader will be referred to \cite{tpejc2018} for the mathematical details. In the same spirit I have kept the references to the minimum; a more extensive list of references can be found in \cite{tpejc2018}.}

\section{Action functionals and their path integrals for a non-relativistic particle}

The dynamical equations governing a classical, non-relativistic  particle  can be obtained from an extremum principle based on \textit{several} kinds of action functionals. Here, I will concentrate   on three different types of action functionals,  of which, two are quite well-known and the third is not so well-known.\footnote{I will only consider a closed, conservative system for which the Hamiltonian is independent of time, leading to the existence of a conserved energy $E$. In fact, most of the discussion will deal with a `free particle' for which the Hamiltonian depends only on the momenta.} 

The first --- most familiar --- one is (i) an action $A_1 [x^\alpha(t)]$ obtained by  integrating a Lagrangian $L_1 = (1/2) m \dot x^\alpha \dot x_\alpha -V(x^\alpha)$ over $dt$. The second approach (ii) uses an action $A_2 [x^\alpha(t), p_\alpha(t)]$ based on a Lagrangian $L_2 = p_\alpha \dot x^\alpha  - H(x^\alpha, p_\alpha)$. In this (Hamiltonian) form of the action we treat $x^\alpha (t) $ and $p_\alpha(t)$ as independent. The third action functional --- which is relatively less familiar --- is (iii) the  Jacobi action functional $A_3[x^\alpha(t)]$ obtained by integrating  $dL_3 = [2m(E-V) dx^\alpha dx_\alpha]^{1/2}$. This action has a square root and is re-parameterization invariant. 
 Varying this action leads to a \textit{spatial} trajectory taken by a particle moving with energy $E$ in a potential $V(x^\alpha)$.  We will have occasion to comment on all these forms of action in the sequel. 
 
 One way to construct the quantum theory of the non-relativistic particle (NRQM) is to postulate that the amplitude $G_{\rm NR}(x_2; x_1)$ for the particle to propagate from the event $x_1 \equiv x^i_1$ to the event $x_2 \equiv x^i_2$ is given by the sum over all paths of the primary amplitude  $\exp(iA)$. Such a sum is ill-defined and needs to be given as a prescription. Different prescriptions --- even when they lead to sensible, calculable results --- will, of course, lead to different theories of NRQM. We know from experiments the kind of quantum dynamics we want to obtain; in fact, we know  the explicit form of $G_{\rm NR}(x_2; x_1)$ from other --- more reliable --- approaches to quantum mechanics in Schrodinger picture or Heisenberg picture. 
 The path integral approach, therefore, needs to be \textit{designed} to reproduce this \textit{particular} form of $G_{\rm NR}(x_2; x_1)$. 
 In principle, one can attempt to do this with any of the valid classical actions, like the set of three actions described above. Let me discuss briefly what happens in each of these cases.
 
 Providing  a prescription for the non-relativistic path integral -- such that it leads to the correct, known form of  $G_{\rm NR}(x_2; x_1)$ --- is a standard textbook exercise as far as the actions $A_1$ and $A_2$ are concerned,
 provided $H(x^\alpha, p_\alpha) $ has the form $H(x^\alpha, p_\alpha) = p_\alpha^2/2m + V(x^\alpha)$. 
  For example, while using $A_1$, the path integral can be defined by slicing the time interval into infinitesimal parts and assuming that, at any given time $t$, the position of the particle $x^\alpha (t)$ is unique; viz., $x^\alpha (t)$ is a single-valued function. \textit{This implies that we only sum over  the paths which always go forward in time $t$. }
  
  A similar procedure can be adopted with the action in $A_2$ as well. But, the set of paths summed over in evaluating $A_2$, compared to evaluating $A_1$, is quite different because we are now treating both $x^\alpha(t)$ and $p_\alpha(t)$ as independent functions; in fact, the physical meaning of such a sum over paths for $A_2$ is not intuitively as clear as the ones used in evaluating $A_1$. (For a further discussion, see Section 4.1 of \cite{tpejc2018}.)
  
  There is yet another subtlety in comparing the approaches based on $A_1$ and $A_2$ in a \textit{more general} context. As long as the Hamiltonian is quadratic in the momenta, one can come up with prescriptions for path integral such that one is led to the same $G_{\rm NR}(x_2; x_1)$ irrespective of whether we use $A_1$ or $A_2$. What is not often emphasized is that this equivalence fails if $H$ does not have a quadratic dependence in $p_\alpha$.  Even in the simple case of a generalized ``free particle'' with $H= H(p_\alpha)$ this equivalence fails.  (The reason for this failure as well as the condition under which the equivalence holds are described in Sections 4.1 and 4.2 of \cite{tpejc2018}.) 
  
  There is a fundamental reason why the definition of path integrals using time slicing works  for $A_1$ and $A_2$. We can verify, using the known functional form of the propagator,  which we are trying to obtain from the path integral, that it satisfies the composition law 
  \begin{equation}
 G_{\rm NR}(x_2, x_1) = \int d^3\bm{x} \ G_{\rm NR}(x_2, x)\,G_{\rm NR}(x, x_1)
   \label{one}
  \end{equation} 
  In fact, the  composition law in \eq{one} is  crucial for consistent interpretation of the Schrodinger wave function when it is propagated by $G_{\rm NR}(x_2; x_1)$. 
  Because $G_{\rm NR}(x_2; x_1)$ obeys this composition law,  the propagator for a finite interval of time can be obtained by iteration of the propagators for  infinitesimal interval of time. A closely related feature is that the action functionals $A_1$ or $A_2$ for an infinitesimal time interval can be used to construct the correct infinitesimal propagator by the rule $G \sim \exp(iA)$  (modulo suitable measure).\footnote{Note that we \textit{know} the functional form of $G_{\rm NR}(x_2; x_1)$  from, say, the Schrodinger equation. The task of the `path-integralist' is to provide a alternative definition for the same. Since we know $G_{\rm NR}(x_2; x_1)$, we can compute it for an infinitesimal time interval; we also know the form of the action for infinitesimal time interval. Therefore we can  directly verify that the rule $G \sim \exp(iA)$ holds; this makes us confident that the time slicing will work, provided the measure can be suitably chosen.} Intuitively speaking, the non-relativistic quantum particle is propagated from one time slice to another  
  by this infinitesimal propagator. 
  
  Consider next the action $A_3$. Because $A_3$ involves a Lagrangian with square root, it is not easy to define a suitable measure for the path integral when we use this action.  However, there are two alternative routes to give meaning to the path integral when we use $A_3$. 
 
 (a) The first approach is to notice that --- after we (somehow) compute the path integral --- the resulting Jacobi propagator $G_J (E; x^\alpha_2, x^\alpha_1)$, should be interpreted as the amplitude for the particle to propagate from the \textit{spatial} location $x^\alpha_1$ to the \textit{spatial} location $x^\alpha_2$
  with energy $E$. From the standard formulation of quantum mechanics,   $G_J (E; x^\alpha_2, x^\alpha_1)$ can be related to the Fourier transform with respect to ($t_2 - t_1$) of the usual propagator $G_{\rm NR}(x_2; x_1)$. Using this fact, one can relate the sum over paths of $A_3$ to the sum over paths of $A_1 $ (or, equivalently, $A_2$) by \textit{postulating} an equivalence. This, of course, is not a very satisfactory approach and can only be thought of as an interpretation of the path integral with $A_3$. 
  
  (b) There exists a   rigorous and natural procedure is to define the path integral involving $A_3$ using  a Euclidean lattice regularization. This approach works   only for a free particle  but that is what we will eventually be concerned with. In the case of a free particle, $A_3$ is proportional to the length $\ell (x^\alpha, x^\alpha_1)$ of the path connecting the two spatial locations and we need to give meaning to the sum over amplitudes of the form $\exp[-\mathcal{P}(E) \ell (x^\alpha, x^\alpha_1)]$ where $\mathcal{P}(E)=\sqrt{2mE}$. This can indeed be defined rigorously in a Euclidean lattice and one can obtain the explicit form of $G_J (E; x^\alpha_2, x^\alpha_1)$. (See Eq. 121 of \cite{tpejc2018}.) An important cross-check for the validity of this procedure is that,  if you Fourier transform $G_J (E; x^\alpha_2, x^\alpha_1)$ --- obtained by the lattice technique --- with respect to $E$, you should get the standard $G_{\rm NR}(x_2; x_1)$. It is easy to verify that this actually happens. Therefore, working in an Euclidean lattice, one can give meaning to $A_3$ for the free particle which is an action with a square root.\footnote{Of course, an arbitrary real function $x^\alpha(t)$ will not map into a real function under analytic continuation to Euclidean time. 
  Therefore, the set of paths summed over, in the lattice regularization method, is not the same as the set of paths summed over while evaluating, say, $A_1$, by time-slicing.}

  Given the facts that:  (i) we have a prescription for evaluating $G_{\rm NR}(x_2; x_1)$ and (ii)  $G_J (E; x^\alpha_2, x^\alpha_1)$ is just a Fourier transform of  $G_{\rm NR}(x_2; x_1)$, it may be possible to provide a direct prescription for evaluating $G_J$ \textit{without} using lattice regularization. (Such a procedure, if it exists, might also work for the Jacobi action for a particle in a potential.) I had attempted to give such a prescription in \cite{pione,pitwo}.  Unfortunately, this prescription does not quite work because of a technical issue with the measure. (For e.g. I have not been able to find a suitable definition of measure such that both equations 1 and 3 of the Appendix in \cite{pione} hold.) This is why I have abandoned this approach in the later works, \cite{tpqft} and \cite{tpejc2018}, in favour of the lattice regularization, which was \textit{also} discussed in detail in \cite{pione,pitwo} --- in Section IIA of \cite{pitwo} and Section 3 of \cite{pione} with  both  sections  having the word ``rigorous'' in their title. I have also emphasized in the later works \cite{tpqft,tpejc2018} that the lattice regularization can be thought of as a \textit{derivation} while other methods involving Jacobi action should be thought of as an \textit{interpretation}.

  \section{Path integrals for a relativistic particle}
  
  A classical relativistic particle can also be  described by different kinds of action functionals somewhat  analogous to the ones we encountered above for the non-relativistic particle. The simplest one ($A_I$) is obtained by integrating  the Lagrangian $L_I = - m (1 - v^2)^{1/2}$ over $dt$. It is also possible to write down the Hamiltonian form of this action $A_{\rm II}$ with $H(p_\alpha) = (p_\alpha p^\alpha + m^2)^{1/2}$. In both these cases, we are considering the paths $x^\alpha (t)$ which are single valued functions of $t$, exactly in the non-relativistic case. It is also possible to work with paths defined as parameterized curves $x^i(\tau)$ and define an action $A_{\rm III}$  containing a non-dynamical variable $N(\tau)$ to ensure reparametrisation invariance. I will discuss the explicit form of this action in the sequel.
  
  Just as in the case of non-relativistic mechanics, one would like to construct the relativistic propagator $G_R(x_2,x_1)$ by  a path integral over any of these actions.
  Again, as in the case of NRQM, you first need to know the answer --- viz. the explicit form of  $G_R(x_2,x_1)$ --- and then you can try to come up with a path integral prescription which will lead to this result. We know from QFT that relativistic particles should be thought of as excitations of an underlying quantum field and the relevant propagator is the expectation value of the time ordered two-point correlation function of the field, usually called the Feynman propagator. This propagator propagates both particles  and anti-particles in a specific manner. Our aim is to obtain such a propagator by defining the path integrals suitably. One can already see the nontriviality in the relativistic case. As is well known, single particle relativistic quantum mechanics does not exist and combining SR with QM  leads to the existence of antiparticles --- which have to be somehow incorporated into the path integral.

Once again, we face the problem that path integrals are intrinsically ill defined and the answer will depend on the prescription  used for defining the path integral.  An important example which illustrates this point is provided by the relativistic action $A_{\rm II}$ in the Hamiltonian form. For \textit{any} system with a Hamiltonian $H= H(p_\alpha)$, it is possible to come up with a natural prescription for evaluating the Hamiltonian form of the path integral. (See Sec. 4.1 of \cite{tpejc2018}.) If you carry out this procedure (which works perfectly well when $H=p^2/2m$) for the relativistic particle with $H= (p_\alpha p^\alpha + m^2)^{1/2}$, you \textit{do get} a well defined propagator.
In fact you get \textit{two} different propagators --- depending on whether you carry out momentum integrations in the time-slices using the measure $d^3p$ (which is not Lorentz invariant) or whether you use the measure $d^3p/2\omega_p$ (which is Lorentz invariant) --- but neither of them is the Feynman propagator we want to obtain. When you use a measure which is not relativistically invariant you get a propagator called Newton-Wigner propagator (see eq. 64 of \cite{tpejc2018}) and when you use a relativistically invariant measure, you get propagator built from just positive frequency modes (see eq. 67 of \cite{tpejc2018}).
  
  The situation is actually worse for the Lagrangian path integral, based on $A_I$ if we use time-slicing.. In the case of Hamiltonians which are quadratic in momenta, the Lagrangian and Hamiltonian forms of the action can be made to give the \textit{same} propagator by a suitable prescription. But in the case of a relativistic particle, the Hamiltonian is not quadratic in momenta and hence the equivalence between path integrals defined using $A_I$ and $A_{\rm II}$ is not guaranteed. As I said before, $A_{\rm II}$ \textit{does } lead to a well defined propagator but it is not the Feynman propagator we want. On the other hand, the most natural prescription for path integral (by time slicing in $t$) using $A_I$ leads to ill-defined, divergent, expression. (See Sec. 4.2 of \cite{tpejc2018}.)
 
 The fact that time slicing in $t$ will not lead to $G_R(x_2; x_1)$ is actually obvious even before we attempt the computation. Recall that the composition law in \eq{one} was crucial for the time slicing to work in the case of non-relativistic particle. The Feynman propagator, which we are trying to obtain from the path integral, does not obey this composition law. Instead, we have the result\footnote{The Feynman propagator obeys a composition law defined through the Klein-Gordan inner product, which is not directly useful for defining the path integral.}
  \begin{equation}
   \int d^4x\ G_R(x_2, x)\, G_R(x,x_1) = i \frac{\partial}{\partial m^2} G_R(x_2; x_1)
  \end{equation} 
  This composition law with an integration over $d^4x$ is difficult to interpret intuitively. But if we use $d^3 x$ the resulting expression in the left hand side will not be Lorentz invariant (and will not obey the composition law either).
  
  An equivalent way of stating this result is the following. In the case of a non-relativistic particle, the infinitesimal form of the propagator (that is, the propagator between $t$ and $t+\epsilon$)  can be expressed in terms of the infinitesimal form of the non-relativistic action $A_1$ in the form $G\sim \exp iA$.  This is not the case for the relativistic action $A_I$. (For more details, see discussion leading to equation (135) in \cite{tpejc2018}.) 
  
  The \textit{conceptual} reason for the failure of time-slicing, which is more intriguing and important, is the following: The time slicing in $t$ leads to sum over paths which only go forward in time. If you use such a prescription and a relativistically invariant measure, it can at best lead only to a propagator containing positive frequency modes --- which is precisely what happens in the Hamiltonian formulation with $A_{\rm II}$. On the other hand, the Feynman propagator which we want to obtain, incorporates both forward and backward propagation.  
  
  So how do we obtain the Feynman propagator from path integral using, say, $A_I$? 
  
  Fortunately, there is a natural and rigorous procedure for doing the same, which works like a miracle. We note that in the Euclidean sector, the action $A_I$ is proportional to the length of the path $\ell (x_2,x_1)$. We can therefore compute the path integral with the Euclidean action in a Euclidean lattice with an appropriate regularization procedure. As demonstrated in Ref. \cite{tpqft,tpejc2018} (as well as in the the earlier works \cite{pione,pitwo}), this procedure indeed leads to the correct Euclidean propagator. Analytically continuing back to the Lorentz sector, one indeed obtains the Feynman propagator.

  As we said before, you cannot obtain the Feynman propagator if you deal with time and space differently and sum over paths which go only forward in time. This is the problem with the time slicing in $t$.  In the Euclidean lattice calculation we treat time and space at equal footing and allow for paths which go both backwards and forwards in time (as well as in space). This ``somehow'' incorporates the existence of antiparticles in the path integral and leads to the Feynman propagator. It is far from clear why this procedure works and I will have more to say about this in the next section.
  
  Are there ways of obtaining the Feynman propagator \textit{without} using the lattice regularization? It is possible to construct some such procedures (see, for e.g., \cite{poly}) but they are less transparent and intuitive compared to the lattice regularization.\footnote{Aside: At first sight, it might appear that the need to renormalize the mass, as a part of lattice regularization,  makes the lattice procedure also somewhat opaque; this is, however, not true and this renormalization should be thought of as part the definition of our prescription. (See, for e.g., Eqn (84) of Ref. \cite{tpejc2018}. This is further clarified in Sec. 5.1 of the same paper.)}  
  I will comment briefly on these procedures for the sake of completeness. 
  
  Since it is crucial to incorporate paths which go back and forth both in time and space, it is convenient to describe the paths in parametric form by the four functions $x^i(\tau)$ where $\tau$ is the parameter. To obtain a suitable prescription, we only have to 
  reverse-engineer the Schwinger proper time representation of the Feynman propagator given by 
  \begin{equation}
 G_R 
=-\frac{i}{16\pi^2}\int_0^\infty\frac{ds}{s^2}\, \exp\left({-im^2 s- \frac{i}{4s}x^2}\right)
\label{gr1}
  \end{equation} 
 We see that the part $s^{-2}\exp (-ix^2/4s)$ can be obtained from a path integral with a quadratic action based on the Lagrangian
 $L_{\rm quad}=-(1/4) (dx^i/d\tau)(dx_i/d\tau)$. (Such a Lagrangian, of course, will give the correct classical equations of motion 
  $(d^2x^i/d\tau^2)=0$ for a massive particle if we identify $\tau$ with the proper time.). The  additional phase $\exp -im^2s$ in \eq{gr1} can be easily included by modifying the Lagrangian to the form $L_{\rm III} = - (1/4) (dx^i/d\tau)(dx_i/d\tau) - m^2$. Starting with an action $A_{\rm III}[x^i(\tau)]$ based on this Lagrangian $L_{III}$,
  treated as a functional of $x^i(\tau)$,
  we can compute the quadratic path integral with time-slicing in $\tau$. We then get  a (Schwinger) propagator $K(x_2,s; x_1,0)$ which could be interpreted as the amplitude for the particle to propagate from $x_1$ to $x_2$, when the parameter describing the evolution varies from $\tau =0$ to $\tau=s$. Since we do not care about the lapse $s$ in the parameter,    it makes sense to integrate over $s$  and \textit{define} the resultant expression as the  the Feynman propagator $G_R(x_2,x_1)$ which we seek. We see from \eq{gr1} that this procedure is guaranteed to give the correct result, which, of course, is a bit of tautology. This approach can be summarized by the relation:
  \begin{equation}
 G_R(x_2;x_1) = \int_{0}^\infty ds\,  K(x_2,s;x_1,0) = \int_{0}^\infty ds\,\sum_{x(\tau)} e^{iA_{III}[x(\tau)]}
   \label{oneoneseven}
  \end{equation} 
  
  The  action based on $L_{\rm III} = - (1/4) (dx^i/d\tau)(dx_i/d\tau) - m^2$, however, has (at least)
   two unsatisfactory features: (a) The specific form of $L_{\rm III}$ requiring the addition of a constant ($-m^2$) is rather ad-hoc. (b) More importantly, the action $A_{\rm III}$  is not re-parameterization invariant under $\tau\to f(\tau)$.  To remedy this, we need to introduce the non-dynamical variable $N(\tau)$ and work with the modified  action 
  \begin{equation}
  A_{\rm III} [N(\tau), x^i(\tau)]\equiv-\int d\tau\ \left[\frac{1}{4N}\, \frac{dx^i}{d\tau}\frac{dx_i}{d\tau} - m^2N\right]
  \label{gia}
  \end{equation}
  We now need to do the functional integral over \textit{both} $x^i(\tau)$ and $N(\tau)$ with suitable measures.\footnote{The result in \eq{oneoneseven} can be thought of as working in $N=1$ gauge. The functional integral over $N$ is interpreted, in this gauge, as ordinary integral over $s$.} It seems possible to go through this exercise  and obtain the Feynman propagator \cite{poly} but the  construction lacks the simplicity of the lattice regularization procedure. 
  
  This approach, incidentally, throws light on another aspect of the relativistic path integral which we alluded to earlier. The quadratic path integral for $K(x_2,s; x_1,0)$ will lead to the standard expression on  slicing the proper time interval into $N$ parts. To get $G_r$ we need to integrate this expression over $s$.
  Before we integrate over $s$, the $K(x_2,s; x_1,0)$ satisfies the standard composition law in \eq{one} in proper time. But the integration over $s$ spoils this feature.  In fact, after integration over $s$, the time-sliced expression will not lead to the exponential of the infinitesimal form of relativistic action. What we get is   expression which attributes to each path  a complicated weightage involving a Bessel function. (See Eq. (135) of \cite{tpejc2018}.)
  As we said before, this is yet another reason for the failure of the straightforward approach which works in the case of NRQM. Moreover, the fact that the weightage for each path is quite unrelated to the action functional makes even time slicing in $\tau$ somewhat suspicious. (See the last part of Sec. 5.2 of \cite{tpejc2018} for further discussion.) It is possible to construct different avatars of this procedure (essentially paraphrasing the approach in \cite{poly}) but  all of them seem to lack the simplicity of lattice regularization. 
  
  \section{Concluding remarks}
  
  I have explained why the lattice regularization procedure leads to an intuitively appealing derivation of the Feynman propagator using a relativistic path integral. This procedure echoes in the calculation of Jacobi propagator for the non-relativistic free particle. The alternative approach using the action in \eq{gia} and defining the path integral with suitable measures for both $N(\tau)$ and $x^i(\tau)$ appears, in comparison, somewhat opaque.  
  
  While the lattice approach leads to the correct result, it raises several technical and conceptual issues which deserve further scrutiny. Among the technical questions, I consider the following two as particularly interesting:
  
  (a) The Jacobian propagator $G_J(E; x^\alpha_2, x^\alpha_1)$ can be evaluated rigorously from the path integral based on Jacobi action using the Euclidean lattice regularization. Similarly, the standard propagator $G_{\rm NR}(x_2, x_1)$ can also be obtained from a path integral defined using the standard time slicing. We also know that the two resulting expressions --- both calculated independently from respective path integrals --- are related by a Fourier transform. This strongly suggests that there should be a way of defining the path integral measure for the Jacobi path integral such that it can be evaluated without lattice regularization. This is indeed possible for a free particle, along the lines of \cite{poly}, but it is not obvious how to do it in the presence of a potential $V(x^\alpha)$. (I attempted this in \cite{pione,pitwo} but, as I mentioned before, there is a technical issue with the measure I used.)
  
 (b) It will be useful to explicitly obtain the non-relativistic propagator from the relativistic one (obtained from, say, $A_{III}$ with integration over $N(\tau)$ and $x^i(\tau)$) paying particular attention to the question of measure in the path integral in Lorentzian sector and its limit. (It is possible to perform the corresponding exercise in the lattice by restricting the paths to those which go only forward in time but this result is rather formal; see sec 5.1 and 5.3 of \cite{tpejc2018}.) This task is  nontrivial because the $c\to\infty$ limit of the Feynman propagator does not lead to the non-relativistic propagator, contrary to one's naive expectation (see Sec. 5.3 of \cite{tpejc2018}). So it is not clear whether the approach based on $A_{III}$ has a natural non-relativistic limit.
  
  I finally come  to the crucial conceptual issue which was the main motivation for me to  pursue this analysis. (This is discussed in detail in Section 8.2 of \cite{tpejc2018}.)  The fact that Feynman propagator can be obtained by a lattice regularization of the relativistic particle (without mentioning at all the notion of a quantum field),  tells us that   we have the algebraic equality:
  \begin{eqnarray}
  \label{oneninetyfour}
 &&\sum_{\rm paths} \exp\left( - \frac{im}{\hbar} \int_1^2 dt\, \sqrt{1-\bm{v}^2} \right)\\
 &&\hskip5em= \theta(t_2-t_1) \bk{0}{A(x_2)A^\dagger(x_1)}{0} + \theta(t_1-t_2) \bk{0}{B(x_1)B^\dagger(x_2)}{0}\nonumber
\end{eqnarray} 
The right hand side is the standard Feynman propagator for a complex scalar field, $\bk{0}{T[\phi(x_2)\phi^\dagger(x_1)]}{0}$, decomposed as 
$\phi(x) = A(x)+B^\dagger(x)$ in terms of two scalar fields $A(x)$ and $B(x)$. Both $A(x)$ and $B(x)$ are (i) built from corresponding creation operators for particles and antiparticles, with positive frequency modes, (ii) satisfy Klein-Gordan equation and (iii) are Lorentz invariant scalars. But neither $A(x)$ not $B(x)$ have vanishing commutators for spacelike separations while $\phi$ does. It is this requirement of relativistic causality that forces us to work with $\phi$ --- which describes both particles and antiparticles at one go --- rather than with $A$ and $B$ separately. The \eq{oneninetyfour} tells us that when $t_2>t_1$, the $A$ type particle propagates forward in time while if $t_2<t_1$ it is the $B$ type particle which propagates forward in time. It is far from obvious how the path integral in the left hand side incorporates both $A$ and $B$ type particles at one go. (Remember that the left hand side, at the face of it,  knows nothing about quantum fields or the existence of   anti-particles.)

In fact the situation is more intriguing. Recall that we can 
 compute the left hand side using lattice regularization in the \textit{Euclidean} lattice. (As far I could see,  it is  not possible to do a corresponding computation in some \textit{spacetime} lattice.) The Euclidean space knows nothing about the light cone structure, time ordering, microcausality etc.  How is it that such a Euclidean computation --- which does not recognize a light cone structure or the notion of causality --- correctly reproduces the Feynman propagator which recognizes all these features? It is  usual in QFT to think of Euclidean continuation (or the Wick rotation) as a mathematical trick to ensure convergence of certain integrals. Equation (\ref{oneninetyfour}) tells us that \textit{this is a very limited, and most probably incorrect, point of view regarding the Euclidean computation}. It certainly captures several vital features in a manner which is not obvious on the surface and this fact deserves further exploration. (I have some specific ideas on this which I hope to describe in a future work.) 
 
 As I said in \cite{tpejc2018}  nobody really understands \eq{oneninetyfour}; that is, nobody has a way of explaining this equality without doing fairly complicated computations. 
  This equality requires a deeper physical understanding which does not exist at present.


\begin{thebibliography}{000}

\bibitem{tpqft}  
T. Padmanabhan, \textit{Quantum Field Theory: The Why, What and How}, Springer, Heidelberg (2016).

\bibitem{tpejc2018} 
T. Padmanabhan, \textit{Eur. Phys. J. C}, \textbf{78}, 563 (2018)  [arXiv:1712.06605].

\bibitem{pione} 
T.Padmanabhan,  \textit{Found.Physics}, \textbf{24}, 1543 (1994).


\bibitem{pitwo}
T. Padmanabhan,  \textit{Phys. Rev. D.}, \textbf{57}, 6206 (1998).

\bibitem{poly} 
A. M. Polyakov, \textit{Gauge Fields and Strings}, Harwood Academic Publishers, Chur, Switzerland;  New York (1987)

\end{thebibliography}
\end{document}